\documentclass[prb,twocolumn,a4paper,superscriptaddress,floatfix,showpacs]{revtex4}
\usepackage[english]{babel}
\usepackage{epsfig}
\usepackage{graphicx}
\usepackage{amsmath,amssymb,amsfonts}
\usepackage{bm}

\usepackage{color}
\usepackage{xcolor}
\usepackage{ulem}

\usepackage{verbatim}

\DeclareMathOperator{\re}{Re}
\DeclareMathOperator{\sign}{sgn}

\begin{document}

\title{
Electronic phase separation in iron pnictides
}

\author{A.~O.~Sboychakov}
\affiliation{Institute for Theoretical and Applied Electrodynamics,
Russian Academy of Sciences,
Moscow, 125412 Russia}
\affiliation{CEMS, RIKEN, Saitama, 351-0198, Japan}

\author{A.~V.~Rozhkov}
\affiliation{Institute for Theoretical and Applied Electrodynamics,
Russian Academy of Sciences,
Moscow, 125412 Russia}
\affiliation{CEMS, RIKEN, Saitama, 351-0198, Japan}

\author{K.~I.~Kugel}
\affiliation{Institute for Theoretical and Applied Electrodynamics,
Russian Academy of Sciences,
Moscow, 125412 Russia}
\affiliation{CEMS, RIKEN, Saitama, 351-0198, Japan}

\author{A.~L.~Rakhmanov}
\affiliation{Institute for Theoretical and Applied Electrodynamics,
Russian Academy of Sciences,
Moscow, 125412 Russia}
\affiliation{CEMS, RIKEN, Saitama, 351-0198, Japan}
\affiliation{Moscow Institute of Physics and Technology,
Dolgoprudnyi, Moscow Region, 141700 Russia}

\author{Franco~Nori}
\affiliation{CEMS, RIKEN, Saitama, 351-0198, Japan}
\affiliation{Department of Physics, University of Michigan, Ann Arbor, MI 48109-1040, USA}

\begin{abstract}
A mechanism for electronic phase separation in iron pnictides is proposed. It is based on the competition between commensurate and incommensurate spin-density-wave phases in a system with an imperfect doping-dependent nesting of a multi-sheeted Fermi surface. We model the Fermi surface by two elliptical electron pockets and three circular hole pockets. The interaction between a charge carrier in a hole band and a carrier in an electron band leads to the formation of spin-density-wave order. The commensurate spin density wave in the parent compound transforms to the incommensurate phase when doping is introduced. We show that, for certain parameter values, the uniform state is unstable with respect to phase separation. The resulting inhomogeneous state consists of regions of commensurate and incommensurate spin-density-wave phases. Our results are in qualitative agreement with recent observations of incommensurate spin density waves and electronic inhomogeneity in iron pnictides.
\end{abstract}



\date{\today}

\maketitle

\section{Introduction}
\label{Intr}

Superconducting iron-based pnictides~\cite{KamiharaJAmChS2008}
attract considerable interest not only due to their high critical
temperatures, but also because of the rich physics of their electron
subsystem. The phase diagram in iron pnictides contains areas of
superconductivity, spin-density wave (SDW) order~\cite{DagottoNatPhys2012}, both commensurate~\cite{dong2008} and incommensurate~\cite{PrattICSDWed}, and even a phase with electronic nematicity.~\cite{KasaharaNature2012}

Moreover, these materials often demonstrate spin and charge inhomogeneity, exhibiting characteristic features of systems with electronic phase separation.~\cite{PSexp1,PSexp2,PSexp3,PSexp4,DagottoNatPhys2012,
DagottoPRB2010phys_reg} The origin of this phase separation is important for understanding the mechanisms driving numerous phase transitions in the phase diagram of iron pnictides.

The phase separation is quite ubiquitous, manifesting itself in different situations where the itinerancy of charge carriers competes with their tendency to localization. The latter is often related to some specific type of magnetic ordering, e.g. antiferromagnetic in manganites or low-spin state in cobaltites. The interplay between the localization-induced lowering of the potential energy and metallicity, providing a gain in the kinetic energy, favors an inhomogeneous ground state, such as nanosize ferromagnetic droplets on an antiferromagnetic insulating background. This type of phase separation has a long history (beginning from the seminal work of Refs.~\onlinecite{NagaevJETPL1967,KasuyaSSC1970}) in the fields of magnetic semiconductors and doped manganites. Electron correlations here also play an important role enhancing the tendency to localization.~\cite{DagottoScience2005} Moreover, it can be shown
that in strongly correlated electron systems with different types
of charge carriers, phase separation can appear even in the
absence of any specific order parameter.~\cite{SboychaPRB2007}

In this paper, we discuss a very different mechanism of phase separation, which is not directly related to electron correlations and could be more relevant in the case of iron pnictides. It is based on the imperfect nesting of different pockets of the Fermi surface. Indeed, it has been known that the SDW ground state in a model with two spherical Fermi surfaces of unequal radius~\cite{RicePRB70} can be unstable with respect to electronic phase separation.~ \cite{tokatly,AAgraphene_doped,OurRicePRB} Variations of the latter model have been used to describe SDW in chromium~\cite{RicePRB70} and graphene bilayer with AA stacking~\cite{AAgraphenePRL}. It is also analogous (but not identical) to the commonly used model of iron pnictides where one deals, roughly speaking, with elliptical rather than spherical electron and hole sheets of the Fermi surface. One can ask if a similar mechanism could apply to pnictides. In this paper, we demonstrate that the answer to this question is positive, and thus the charge inhomogeneity can result from a purely electronic mechanism. This finding is important for the interpretation of experimental data on charge inhomogeneity and for understanding the nature of the coexistence of the order parameters in iron pnictides.

For definiteness, we will focus here on iron pnictides, although iron chalcogenides also exhibit phase separation.~\cite{Bianconi1,Bianconi2} However, the physics related to the Fermi surface nesting may not be directly applicable to iron chalcogenides, especially to those containing alkaline atoms. Indeed, some chalcogenides do not have hole pockets, but nevertheless exhibit antiferromagnetism with rather high N\'eel temperatures.~\cite{chalcWang,chalcZhang,chalcMou}
Moreover, the electron correlation effects in chalcogenides seem to be more pronounced than in pnictides.~\cite{DagottoNatPhys2012}
Therefore, the case of iron chalcogenides requires a separate
consideration.

This paper is organized as follows. In Sec.~\ref{model} we discuss the choice of the model Hamiltonian. Section~\ref{sdw} deals with the study of homogeneous SDW order in the mean-field approximation. The instability of the homogeneous state is proved in Sec.~\ref{separation}, where the phase diagram is constructed as well. The results are discussed in Sec.~\ref{discussion}. In Sec.~\ref{Concl}, we summarize the main conclusions of the paper, list the assumptions used, and formulate possible tasks for future work.

\section{Model}\label{model}

\subsection{Kinetic energy}

\begin{figure}[t!]\centering
    \includegraphics[width=0.98\columnwidth]{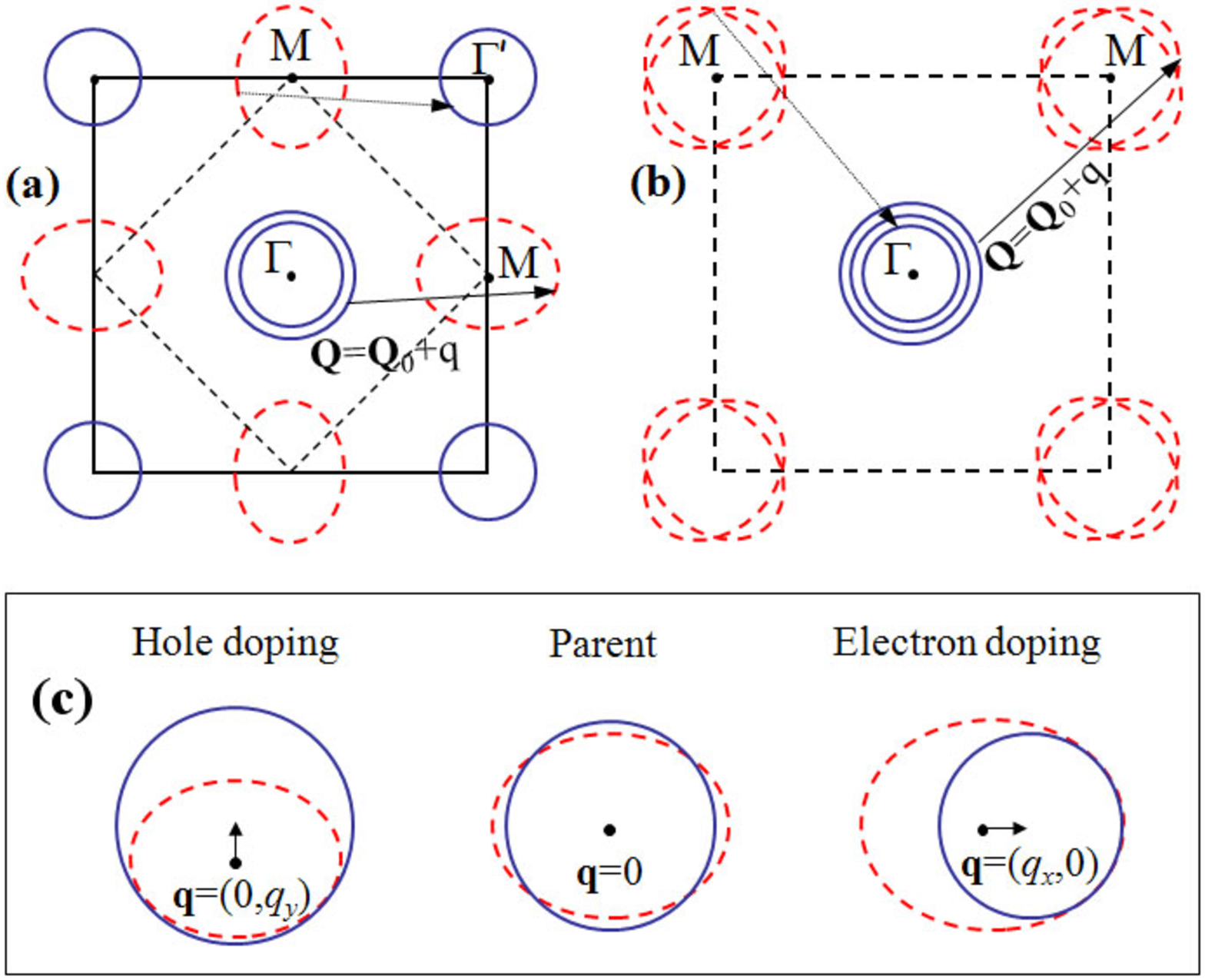}
\caption{(Color online) Schematic illustration of the Fermi surface of iron pnictides in the unfolded (a) and folded (b) Brillouin zone (BZ). We have three hole pockets located near the center of the folded BZ [shown by (blue) solid curves]. In the unfolded BZ, one of the hole pockets moves to the corner of the BZ. For the unfolded BZ (a), the electron pockets are elliptic [(red) dashed curves] and located near the $(0,\pi/a)$ and $(\pi/a,0)$ points, while in the folded BZ, they are represented by overlapping ellipses located at the corners. Arrows show the possible nesting vectors between hole an electron pockets, which give rise to the SDW order. (c) Illustration of the possible nesting at different doping levels.}
\label{FermiSurface}
\end{figure}

Unlike cuprates, which are believed to be in the strong electron-electron interaction regime, iron pnictides may be described by a weak-interaction model (see, e.g., the discussion in Sec.~IIIA of Ref.~\onlinecite{stewart_rmp2011}). In this approach, the shape of the Fermi surface and the value of the Fermi velocity are the only relevant single-electron band parameters. In the literature, the Fermi surface of iron pnictides is typically described using two related approaches, which we briefly describe below. In Fig.~\ref{FermiSurface}(a), we plot the Fermi surface within the so called unfolded Brillouin zone,~\cite{stewart_rmp2011,RichardRPrPh201,RaghuPRB2008two-orb,
DaghoferPRL2008two-orb,DaghoferPRL2010three-orb,HuHaoPRX2012,ErChub,Calderon,Fink} which corresponds to the square lattice of iron atoms, with one Fe atom per unit cell and lattice constant $a$.  In this representation, two quasi-two-dimensional nearly-circular hole pockets are centered at the $\Gamma (0,0)$ point, one more circular hole pocket is located near the $\Gamma'\,(\pi,\,\pi)$ point, and two elliptical-shaped electron pockets are centered at the ${\rm M} (0,\pi/a)$ and ${\rm M} (\pi/a,0)$ points [see Fig.~\ref{FermiSurface}(a)]. At the same time, the actual unit cell of pnictides contains two Fe atoms since in the crystal lattice the pnictogen atoms are located in non-equivalent positions. The (folded) Brillouin zone corresponding to this unit cell is obtained by folding the Brillouin zone shown in Fig.~\ref{FermiSurface}(a) by dashed lines and consequent rotation by $45^{\circ}$. The Fermi surface in the folded Brillouin zone is shown in Fig.~\ref{FermiSurface}(b). In this figure, all three hole pockets are situated near the $\Gamma (0,0)$ point, while electron pockets represented by overlapping ellipses are located near the ${\rm M} (\pm\pi/\bar{a},\pi/\bar{a})$ points, with $\bar{a}=a\sqrt{2}$. Formulating our model, we will make several simplifications. First, we neglect the effects associated with non-equivalent positions of the Fe atoms. Consequently, the use of unfolded Brillouin zone is sufficient. Second, we will neglect the 3D structure of the material and study only the 2D model.

Thus, the Hamiltonian of the model has the form
\begin{eqnarray}
H = H_0 + H_{\rm int},
\end{eqnarray}
where the kinetic energy term $H_0$ is given by
\begin{eqnarray}
\label{H0}
H_0
=
\sum_{\mathbf{k}\lambda\sigma}
\varepsilon^{h}_{\lambda\mathbf{k}}
a^{\dag}_{\mathbf{k}\lambda\sigma}
a^{\phantom{\dag}}_{\mathbf{k}\lambda\sigma}+
\sum_{\mathbf{k}s\sigma}
\varepsilon^{e}_{s\mathbf{k}}
b^{\dag}_{\mathbf{k}s\sigma}b^{\phantom{\dag}}_{\mathbf{k}s\sigma}\,.
\end{eqnarray}
In this equation, $a^{\dag}_{\mathbf{k}\lambda\sigma}$,
$a^{\phantom{\dag}}_{\mathbf{k}\lambda\sigma}$
($b^{\dag}_{\mathbf{k}s\sigma}$, $b^{\phantom{\dag}}_{\mathbf{k}s\sigma}$) are the creation and annihilation operators for electrons in the hole-like (electron-like) bands $\lambda=1,2,3$ ($s=1,2$) with  spectra $\varepsilon^{h}_{\lambda\mathbf{k}}$  ($\varepsilon^{e}_{s\mathbf{k}}$).

To simplify our formalism, we assume that near the Fermi level the bands have quadratic dispersions. For circular hole-like bands we have ($\hbar = 1$)
\begin{eqnarray}
\label{Spec_hole}
\varepsilon^{h}_{1\mathbf{k}}&=&-\frac{v^{h}_{F}k^2}{2k_F}+
\frac{v^{h}_{F}k_F}{2}-\mu, \\
\varepsilon^{h}_{2\mathbf{k}}
&=&-\frac{v^{h}_{2F}k^2}{2k_F}+\frac{v^{h}_{2F}k_F}{2}
-\Delta\epsilon_{2}-\mu\,,\label{Spec_hole2}\\
\varepsilon^{h}_{3\mathbf{k}+\bar{\mathbf{Q}}}
&=&-\frac{v^{h}_{3F}k^2}{2k_F}+\frac{v^{h}_{3}k_F}{2}
-\Delta\epsilon_{3}-\mu\,,\label{Spec_hole3}
\end{eqnarray}
where $k_F$ is the Fermi momentum, $\mu$ is the chemical potential,
$v^{h}_{F}$ and $v^{h}_{2,3F}$ are the Fermi velocities for the hole bands, and $\bar{\mathbf{Q}}=(\pi/a,\pi/a)$. The energy shifts $\Delta\epsilon_{2,3}$ in Eqs.~\eqref{Spec_hole2} and~\eqref{Spec_hole3} determine the difference in radii of the hole pockets: for $\Delta\epsilon_{i}>0$ the radius of the hole pocket $i=2,3$ is smaller than that of the hole pocket $1$.

As mentioned above, the electron components of the Fermi surface are
elliptic.  For these, the dispersion near the Fermi surface is given by the following relations
\begin{eqnarray}
\label{Spec_electron1}
\varepsilon^e_{1{\bf k + Q_0}}
=
\varepsilon (k_x, k_y),
\quad
\varepsilon^e_{2{\bf k + Q_0'}}
=
\varepsilon (k_y, k_x),
\end{eqnarray}
where the function $\varepsilon$ is equal to
\begin{eqnarray}
\varepsilon (k_x, k_y)
=
\frac{v^{e}_{F}k^2}{2k_F}
-\frac{v^{e}_{F}k_F}{2}+
\frac{\alpha v^{e}_{F}}{k_F}\left(k_y^2-k_x^2\right)-\mu\, .
\end{eqnarray}
The centers of the elliptic bands are
$\mathbf{Q}_0=(\pi/a,0)$ and $\mathbf{Q}'_0=(0,\pi/a)$,
and $v^{e}_{F}$ is the Fermi velocity for the electron bands averaged over the Fermi surface. Note that in general $v^{h}_{F}\neq v^{h}_{2F}\neq v^{h}_{3F}\neq v^{e}_{F}$. The parameter $\alpha$ defines the ellipticity of the electron pockets. For the Fermi surface structure corresponding to Fig.~\ref{FermiSurface}(a,b), $\alpha$ is positive. In this case, the major axes of the ellipses are directed toward the $\Gamma$ point. For $\alpha<0$, the ellipses are rotated by $90^{\circ}$ around their centers.

\subsection{Interaction Hamiltonian}\label{interaction_ham}

Due to the multi-sheeted structure of the Fermi surface, the interaction Hamiltonian $H_{\rm int}$, in general, must include a number of terms describing interactions between charge carriers in different bands. However, since we are interested in the SDW
order, most of these terms may be omitted for they do not contribute to SDW phase transition \cite{RicePRB70}. For this reason, we ignore the electron-electron and hole-hole interactions.

It is known~\cite{DagottoNatPhys2012} that magnetic order in pnictides is ``stripy". That is, the value of local magnetic moment oscillates along one of the crystal axes, remaining constant along the other axis. In the general case, each hole and electron bands are coupled to each other. However, to reproduce the striped magnetic structure it is sufficient to couple one electron and one hole band.~\cite{ErChub} For example, the SDW with oscillations along the $x$-direction occurs if we couple electrons in the hole band $1$ and electron band $1$, or electrons in the hole band $3$ and electron band $2$ [see the solid and dotted arrows in Fig.~\ref{FermiSurface}(a,b)]. In both cases, the nesting vectors will be nearly the same and equal approximately (exactly, for the commensurate SDW state) to $\mathbf{Q}_0=(\pi/a,0)$. Here we assume that only one hole band and one electron band take part in the formation of SDW order. For definiteness, assume that these are the hole band $\varepsilon^h_1$ and the electron band $\varepsilon^e_1$ (although our results remain the same for any pair of hole and electron bands). Consequently, only the interaction between $\varepsilon^h_1$ and $\varepsilon^e_1$ is crucial for the stabilization of the ordered phase. All other interaction terms will be discarded. As we will show below, even this simplified model allows to explain the experimental data on the commensurate and incommensurate SDW order as well as the phase separation. How our results change if we go beyond this simplification, will be discussed in Section~\ref{discussion}.

Keeping these considerations in mind, we split the model Hamiltonian into two parts, magnetic, $H_{\rm m}$, and nonmagnetic (``reservoir''), $H_{\rm r}$
\begin{eqnarray}
H&=&H_{\rm m}+H_{\rm r}\,,\label{Hmr}\\
H_{\rm m}&=&
\sum_{\mathbf{k}\sigma}
\left[
	\varepsilon^{h}_{1\mathbf{k}}
	a^{\dag}_{\mathbf{k}1\sigma}
	a^{\phantom{\dag}}_{\mathbf{k}1\sigma}
	+
	\varepsilon^{e}_{1\mathbf{k}}
	b^{\dag}_{\mathbf{k}1\sigma}
	b^{\phantom{\dag}}_{\mathbf{k}1\sigma}
\right]
\nonumber
\\
&&
+\frac{V_1}{\cal N}
\!\!\sum_{\mathbf{kk}'\mathbf{K}\sigma\sigma'}
\!\!\!
a^{\dag}_{\mathbf{k}+\mathbf{K}1\sigma}
a^{\phantom{\dag}}_{\mathbf{k}1\sigma}
b^{\dag}_{\mathbf{k}'-\mathbf{K}1\sigma'}
b^{\phantom{\dag}}_{\mathbf{k}'1\sigma'}\,,
\label{Hm}\\
H_{\rm r}&=&
\sum_{\mathbf{k}\sigma}
\left[
	 \sum_{\lambda=2,3}\varepsilon^{h}_{\lambda\mathbf{k}}
	a^{\dag}_{\mathbf{k}\lambda\sigma}
	a^{\phantom{\dag}}_{\mathbf{k}\lambda\sigma}
	+
	\varepsilon^{e}_{2\mathbf{k}}
	b^{\dag}_{\mathbf{k}2\sigma}
	b^{\phantom{\dag}}_{\mathbf{k}2\sigma}
\right]\!.
\label{Hr}
\end{eqnarray}
Here ${\cal N}$ is the number of Fe atoms in a layer, $V_1>0$ is the coupling constant characterizing the Coulomb interaction between the bands $\varepsilon^{h}_{1\mathbf{k}}$ and $\varepsilon^{e}_{1\mathbf{k}}$. Below we will refer to these bands as magnetic bands. The bands $\varepsilon^{h}_{2\mathbf{k}}$, $\varepsilon^{h}_{3\mathbf{k}}$, and and $\varepsilon^{e}_{2\mathbf{k}}$ will be called nonmagnetic, since in our model, they do not contribute to the magnetic order parameter.

Model~\eqref{Hmr} is a generalization of the Rice model, proposed in
Ref.~\onlinecite{RicePRB70} for the description of the incommensurate SDW order in chromium. The Hamiltonian of Ref.~\onlinecite{RicePRB70} has two bands, which participate in the magnetic transition, and a
``reservoir" (nonmagnetic bands corresponding to $H_{\rm r}$, where the subscript `r' stand for ``reservoir"). Unlike Eq.~(\ref{Hmr}), the bands responsible for magnetic ordering in the Rice model have a spherical Fermi surface; therefore, at certain filling, the nesting is perfect.

\section{Incommensurate SDW order}
\label{sdw}

\subsection{Mean-field equations}

We now consider Hamiltonian~\eqref{Hmr} in the mean-field approximation. As we pointed out above, the Coulomb interaction in iron pnictides is weak, therefore, we assume below that
\begin{equation}
\label{limitV}
V_1/\varepsilon_F\ll1\,,
\end{equation}
where the Fermi energy $\varepsilon_F$ is defined as
\begin{equation}\label{EF}
\varepsilon_F=\frac{v^e_F+v^h_F}{2}k_F\equiv v_Fk_F\,.
\end{equation}
The weak-coupling condition
Eq.~(\ref{limitV})
guarantees the applicability of the mean-field approximation.

We will study the stability of the following SDW order parameter:
\begin{equation}
\label{DeltaQ}
\Delta=\frac{V_1}{\cal N}\!\sum_{\mathbf{k}}
\left\langle
	a^{\dag}_{\mathbf{k}1\uparrow}	
	b^{\phantom{\dag}}_{\mathbf{k}+\mathbf{Q}1\downarrow}
\right\rangle
=
\frac{V_1}{\cal N}
\!\sum_{\mathbf{k}}
\left\langle
	a^{\dag}_{\mathbf{k}-\mathbf{q}1\uparrow}
	b^{\phantom{\dag}}_{\mathbf{k}+\mathbf{Q}_01\downarrow}
\right\rangle,
\end{equation}
where the nesting vector ${\bf Q}$ is equal to
\begin{eqnarray}
\mathbf{Q}=\mathbf{Q}_0+\mathbf{q}.
\end{eqnarray}
When ${\bf q} = 0$, our SDW is commensurate, whereas if ${\bf q}$ is small, but non-zero, it is incommensurate. Other types of order parameter will be discussed in Sec.~\ref{discussion}.

Note here that different parts of each pocket of the Fermi surface have different orbital composition.~\cite{RichardRPrPh201,ScalapinoPRB2010,Suzuki2011}
The pairing is more favorable between those parts of the Fermi surfaces which have similar orbital character. Within our approach,  such feature can be accounted by introducing a momentum-dependent coupling $V_1$. Such a generalization would lead to a pronounced wave-vector dependence of the SDW order parameter, and could affect
the transition to the SDW state and the structure of the incommensurate SDW phase. However, the reliable calculation of the momentum-dependent $V_1$ is a complicated task going beyond the scope of the present study. Furthermore, unless the variation of $V_1$ is extremely strong, they do not affect phase separation qualitatively. Thus, we choose to work with a momentum-independent coupling constant.

The magnetization corresponding to the SDW order parameter $\Delta$ lies in $xy$ plane. For the commensurate SDW case, in real space the ``stripy" order is observed. Namely, the magnetization direction remains constant when one moves along the direction normal to
${\bf Q}_0$. However, when one moves parallel to ${\bf Q}_0$,
the magnetization reverses its direction from one iron atom to the next iron atom. For incommensurate SDW, this ``stripy" pattern slowly rotates in the $xy$ plane: the local rotation angle $\phi ({\bf R})$ at the point ${\bf R}$ is equal to $({\bf q} {\bf R})$
[see panel (c) and (d) in Fig.~\ref{FigDeltaQ}].

In the mean-field approximation, the magnetic Hamiltonian
Eq.~\eqref{Hm} takes the form
\begin{eqnarray}
\label{HmMF}
H_{\rm m}^{\rm MF}
\!\!\!\!&=&\!\!\!\!
\sum_{\mathbf{k}\sigma}\!
\left[
	\varepsilon^{h}_{1\mathbf{k}-\mathbf{q}}
	c^{\dag}_{\mathbf{k}\sigma}
	c^{\phantom{\dag}}_{\mathbf{k}\sigma}
	\!\!+
	\varepsilon^{e}_{1\mathbf{k}+\mathbf{Q}_0}
	d^{\dag}_{\mathbf{k}\sigma}
	d^{\phantom{\dag}}_{\mathbf{k}\sigma}
\right.
\nonumber
\\
&&
-\left.
	\Delta\left(
		c^{\dag}_{\mathbf{k}\sigma}
		d^{\phantom{\dag}}_{\mathbf{k}-\sigma}
		+
		d^{\dag}_{\mathbf{k}\sigma}
		c^{\phantom{\dag}}_{\mathbf{k}-\sigma}
	\right)
	+\Delta^2/V_1
\right],
\end{eqnarray}
where we introduce the new operators
$c^{\dag}_{\mathbf{k}\sigma}=a^{\dag}_{\mathbf{k}-\mathbf{q}1\sigma}$,
$d^{\dag}_{\mathbf{k}\sigma}=b^{\dag}_{\mathbf{k}+\mathbf{Q}_01\sigma}$,
and $\Delta$ is assumed to be real. The Hamiltonian
Eq.~\eqref{HmMF}
can be easily diagonalized. The quasiparticle energies are
\begin{eqnarray}
\label{SpecD}
E^{(1,2)}_{\mathbf{k}}
&=&
\frac{
	\varepsilon^{e}_{1\mathbf{k}+\mathbf{Q}_0}
	+
	\varepsilon^{h}_{1\mathbf{k}-\mathbf{q}}
     }{2}\mp
\nonumber
\\
&&
\sqrt{
	\Delta^2
	+
	\frac{1}{4}
	\left(
		\varepsilon^{e}_{1\mathbf{k}+\mathbf{Q}_0}
		-
		\varepsilon^{h}_{1\mathbf{k}-\mathbf{q}}
	\right)^2
     }\,.
\end{eqnarray}
The grand potential $\Omega_{\rm m}$ (per one Fe atom) corresponding to $H_{\rm m}^{\rm MF}$ is ($k_B = 1$)
\begin{equation}
\label{OmegaM}
\Omega_{\rm m}
=
-2T\sum_{s}
\int\limits_{BZ}\!\!
\frac{
	v_0\;d^2\mathbf{k}
     }
     {
	(2\pi)^2
     }
\ln\left[
	1+e^{-E^{(s)}_{\mathbf{k}}/T}
  \right]
+
\frac{2\Delta^2}{V_1}\,,
\end{equation}
where $v_0$ is the volume of the unit cell, and the integration is performed over the 2D Brillouin zone. The SDW gap $\Delta$
and the nesting vector $\mathbf{Q}$ are found from the minimization of $\Omega_{\rm m}$
\begin{equation}
\label{EqDeltaQ0}
\frac{\partial\Omega_{\rm m}}{\partial\Delta}=0\,,\;\;\frac{\partial\Omega_{\rm m}}{\partial\mathbf{q}}=0\,.
\end{equation}

It will be shown below that in the weak-coupling limit
Eq.~(\ref{limitV}), the SDW order can exist only if the deviation from the perfect nesting is small, that is:
\begin{equation}\label{limits}
|\alpha|\ll1\,.
\end{equation}
In addition, the gap $\Delta$ is small compared to $\varepsilon_F$, and the deviation ${\bf q}$ of the nesting vector $\mathbf{Q}$
from the commensurate value $\mathbf{Q}_0$ is also small
\begin{equation}\label{limitQ}
|\mathbf{q}|
=
|{\bf Q} - {\bf Q}_0|
\sim
\Delta/v_F\ll k_F.
\end{equation}

Restricting ourselves to the limit of zero temperature, we can write the first of Eqs.~\eqref{EqDeltaQ0}
in the following form
\begin{equation}
\label{EqDelta}
1=\frac{V_1}{2}\int\!\!\frac{v_0\,d^2\mathbf{k}}
{(2\pi)^2}\frac{1-\Theta\left(E^{(1)}_{\mathbf{k}}
\right)-\Theta\left(-E^{(2)}_{\mathbf{k}}\right)}
{\sqrt{\Delta^2+\displaystyle\frac{1}{4}
\left(\varepsilon^{e}_{1\mathbf{k}+\mathbf{Q}_0}
-\varepsilon^{h}_{1\mathbf{k}-\mathbf{q}}\right)^2}}\,,
\end{equation}
where $\Theta(x)$ is the step function.

When the lower band $E^{(1)}_{\mathbf{k}}$ is filled, while the upper band $E^{(2)}_{\mathbf{k}}$ is empty, the gap attains its maximum value $\Delta_0$. In this case, $\Theta(E^{(1)}_{\mathbf{k}})=\Theta(-E^{(2)}_{\mathbf{k}})=0$
for any $\mathbf{k}$, and Eq.~\eqref{EqDelta} becomes
\begin{eqnarray}
\label{EqDelta0}
1=\frac{V_1}{2}\int\!\!\!dE\,\frac{\bar{\rho}(E)}{\sqrt{\Delta_0^2+E^2}},
\end{eqnarray}
where the generalized density of states is defined as
\begin{eqnarray}
\label{Rho}
\bar{\rho}(E)=\int\!\!\frac{v_0\;d^2\mathbf{k}}{(2\pi)^2}\,
\delta\!\!\left(E-\frac{\varepsilon^{e}_{1\mathbf{k}
+\mathbf{Q}_0}-\varepsilon^{h}_{1\mathbf{k}}}{2}\right)\,.
\end{eqnarray}
Evaluating $\bar{\rho}(E)$, we set $\mathbf{q}=0$, since taking into account values of the order of $|\mathbf{q}|\sim\Delta/v_F$
gives only second-order corrections in Eq.~\eqref{EqDelta0}.

If $|E|\ll\varepsilon_F$, the function $\bar{\rho}(E)$ can be calculated explicitly using Eqs.~\eqref{Spec_hole} and~\eqref{Spec_electron1} for the band spectra
$\varepsilon^{e}_{1\mathbf{k}+\mathbf{Q}_0}$ and
$\varepsilon^{h}_{1\mathbf{k}}$. As a result, near the Fermi surface we obtain
\begin{equation}
\label{Rho0}
\bar{\rho}(E) \approx \bar{\rho}(0)
=
\frac{v_0 k_F^2}{2\pi\varepsilon_F}\,,\;\;\;\;\;|E|\ll\varepsilon_F\,.
\end{equation}
When the energy $E$ is of the order of the band width,
$\bar{\rho}(E)$ vanishes. This makes the integral in
Eq.~(\ref{EqDelta0}) convergent, and one can derive the usual BCS-like expression for the gap
\begin{equation}
\label{Delta0est}
\Delta_0
\approx
\varepsilon_F
\exp\!\left(\!
		-\frac{2\pi\varepsilon_F}{v_0 k_F^2V_1}
    \right)\,.
\end{equation}

If the sample is doped, then $E^{(1)}_{\mathbf{k}}>0$ or
$E^{(2)}_{\mathbf{k}}<0$, for some range of $\mathbf{k}$, and the equation for the band gap becomes
\begin{equation}
\label{EqDeltaX}
\ln\frac{\Delta_0}{\Delta}
=
\int\!\!\frac{\varepsilon_F\, d^2\mathbf{k}}{4\pi k_F^2}
\frac{
	\Theta\left(E^{(1)}_{\mathbf{k}}\right)
	+
	\Theta\left(-E^{(2)}_{\mathbf{k}}\right)}
   {\sqrt{\Delta^2+\displaystyle\frac{1}{4}\left(
			\varepsilon^{e}_{1\mathbf{k}+\mathbf{Q}_0}-
			\varepsilon^{h}_{1\mathbf{k}-\mathbf{q}}\right)^2}}\,.
\end{equation}
Substituting Eqs.~\eqref{Spec_hole} and \eqref{Spec_electron1}
into Eq.~\eqref{EqDeltaX}, and taking into account Eqs.~\eqref{limits} and~\eqref{limitQ}, after straightforward algebra we derive the equation for the band gap in the form
\begin{eqnarray}
\label{EqDeltaFin}
\ln\frac{1}{\delta}
=
\int\limits_{0}^{2\pi}\!\!
\frac{d\varphi}{2\pi}
\re \left\{
		{\rm cosh}^{-1}
		\left[
			\frac{\nu_0(\mathbf{p},\varphi)-\nu}{\delta}
		\right]
     \right\}\,,
\end{eqnarray}
where,
\begin{eqnarray}
\label{nu0}
\nu_0(\mathbf{p},\varphi)
&=&
p_x\cos\varphi+p_y\sin\varphi-\frac{\bar{\alpha}}{2}\cos2\varphi,\;\;
\\
\bar{\alpha}&=&\frac{\alpha\,\varkappa\,\varepsilon_F}{\Delta_0},
\label{alpha}
\end{eqnarray}
and we introduce the following dimensionless quantities
\begin{equation}\label{defs}
\varkappa=\frac{2\sqrt{v_F^ev_F^h}}{v_F^e+v_F^h},\;\;
\delta=\frac{\Delta}{\Delta_0},\;\;\nu=\frac{\mu}{\varkappa\Delta_0},\;\;
\mathbf{p}=\frac{\varkappa v_F\mathbf{q}}{2\Delta_0}\,.
\end{equation}

Transforming similarly the second of Eqs.~\eqref{EqDeltaQ0}, we obtain the equation for the nesting vector
$\mathbf{Q} = \mathbf{Q}_0+2\Delta_0\mathbf{p}/v_F\varkappa$,
\begin{eqnarray}
\label{EqQ}
\left(\begin{array}{c}\!\!p_x\!\!\\\!\!p_y\!\!
\end{array}\right)&=&\int\limits_{0}^{2\pi}\!\!\frac{d\varphi}{\pi}
\left(\begin{array}{c}\!\!\cos\varphi\!\!\\\!\!\sin\varphi\!\!
\end{array}\right)\sign\left(\nu_0(\mathbf{p},\varphi)-\nu\right)\times\nonumber\\
&&\re\sqrt{\left(\nu_0(\mathbf{p},\varphi)-\nu\right)^2-\delta^2}\,.
\end{eqnarray}

Equations~\eqref{EqDeltaFin} and~\eqref{EqQ} determine the SDW band gap $\Delta$ and the nesting vector $\mathbf{Q}$ as functions of $\mu$. However, experiments are performed at fixed doping, not chemical potential. Thus, we have to relate the electron density and $\mu$. The total number of electrons per iron atom, $n(\mu)$, is the sum of the number of electrons in the nonmagnetic and the magnetic bands
$n(\mu) = n_{\rm r}(\mu) + n_{\rm m} (\mu)$, where
\begin{eqnarray}
\nonumber
n_{\rm r}(\mu)
&=&
\frac{2}{\cal N}
\sum_{\mathbf{k}}
\left[
	 \sum_{\lambda=2,3}\Theta(-\varepsilon^{h}_{\lambda\mathbf{k}})
	+
	\Theta(-\varepsilon^{e}_{2\mathbf{k}})
\right],\\
n_{\rm m}(\mu)&=&-\frac{\partial\Omega_{\rm m}}{\partial\mu}.
\end{eqnarray}
The doping is defined by
\begin{eqnarray}
x(\mu)=n(\mu)-n(0).
\end{eqnarray}
Performing calculations similar to those described above, we obtain for the doping level
\begin{eqnarray}\label{EqX}
\frac{x}{x_0}&=&r\nu-\int\limits_{0}^{2\pi}\!\!\frac{d\varphi}{2\pi}\,
\sign\left(\nu_0(\mathbf{p},\varphi)-\nu\right)\times\nonumber\\
&&\re\sqrt{\left(\nu_0(\mathbf{p},\varphi)-\nu\right)^2-\delta^2}\,,
\end{eqnarray}
where
\begin{equation}\label{defs1}
x_0=\frac{2v_0k_F^2}{\pi\varkappa}\frac{\Delta_0}{\varepsilon_F},\;\;
r=\frac{(v^{h}_{2F})^{-1}+(v^{h}_{3F})^{-1}+
(v^{e}_{F})^{-1}}{(v^{h}_{F})^{-1}+(v^{e}_{F})^{-1}}\,.
\end{equation}
The first (second) term in Eq.~\eqref{EqX} is the nonmagnetic (magnetic) contribution $x_{\rm r}$ ($x_{\rm m}$) to the total doping $x$. The parameters $x_0$ and $r$ in Eq.~\eqref{defs1} have clear physical meanings: $x_0$ defines the characteristic scale of doping where SDW order exists, while $r$ is the ratio of the densities of states of electrons in the non-magnetic and magnetic bands.
Equations~\eqref{EqDeltaFin},~\eqref{EqQ}, and~\eqref{EqX} form a closed system of equations for the self-consistent determination of
$\Delta(x)$, $\mathbf{Q}(x)$, and $\mu(x)$.

\subsection{Results: Homogeneous state}
\label{results}

Our numerical analysis reveals that
Eq.~\eqref{EqDeltaFin} has no solutions if
$|\bar{\alpha}|>2.0$. Thus, the equality
$|\bar{\alpha}|=2.0$ determines the critical value of the ellipticity parameter $\alpha_c$: for a given coupling constant $V_1$, the SDW ordering occurs only if
\begin{eqnarray}
\label{alpha_c}
|\alpha| < \alpha_c = \frac{2 \Delta_0}{\varkappa \varepsilon_F}.
\end{eqnarray}
Conversely, this
condition~Eq.~(\ref{alpha_c})
may be re-formulated as a requirement on the
interaction strength: for given band parameters ($\alpha$, $\varkappa$,
etc.) the SDW is stable only if
\begin{equation}
\label{Vc}
V_1 > V_c=\frac{2\pi \varepsilon_F}{\displaystyle v_0 k_F^2
\ln \left(\frac{2}{|\alpha|\varkappa}\right)} \,.
\end{equation}

It is seen from Eqs.~\eqref{defs} and~\eqref{defs1} that the parameter $\varkappa$ only renormalizes the dimensionless quantities $\nu$, $\mathbf{p}$, and $x_0$. Thus, at fixed $\bar{\alpha}$ and $r$, it is enough to know the functions $\Delta(x)$, $\mathbf{q}(x)$, and $\mu(x)$ for $\varkappa=1$. For $\varkappa\neq1$ these functions are found by simple rescaling as $\Delta(x\varkappa)$, $\mathbf{q}(x\varkappa)/\varkappa$, and $\varkappa\mu(x\varkappa)$. For this reason, below we study only the $\varkappa=1$ case. Note that, typically, for a real material all Fermi  velocities are of the same order, thus $\varkappa \sim 1$.

\begin{figure}[t]\centering
    \includegraphics[width=0.98\columnwidth]{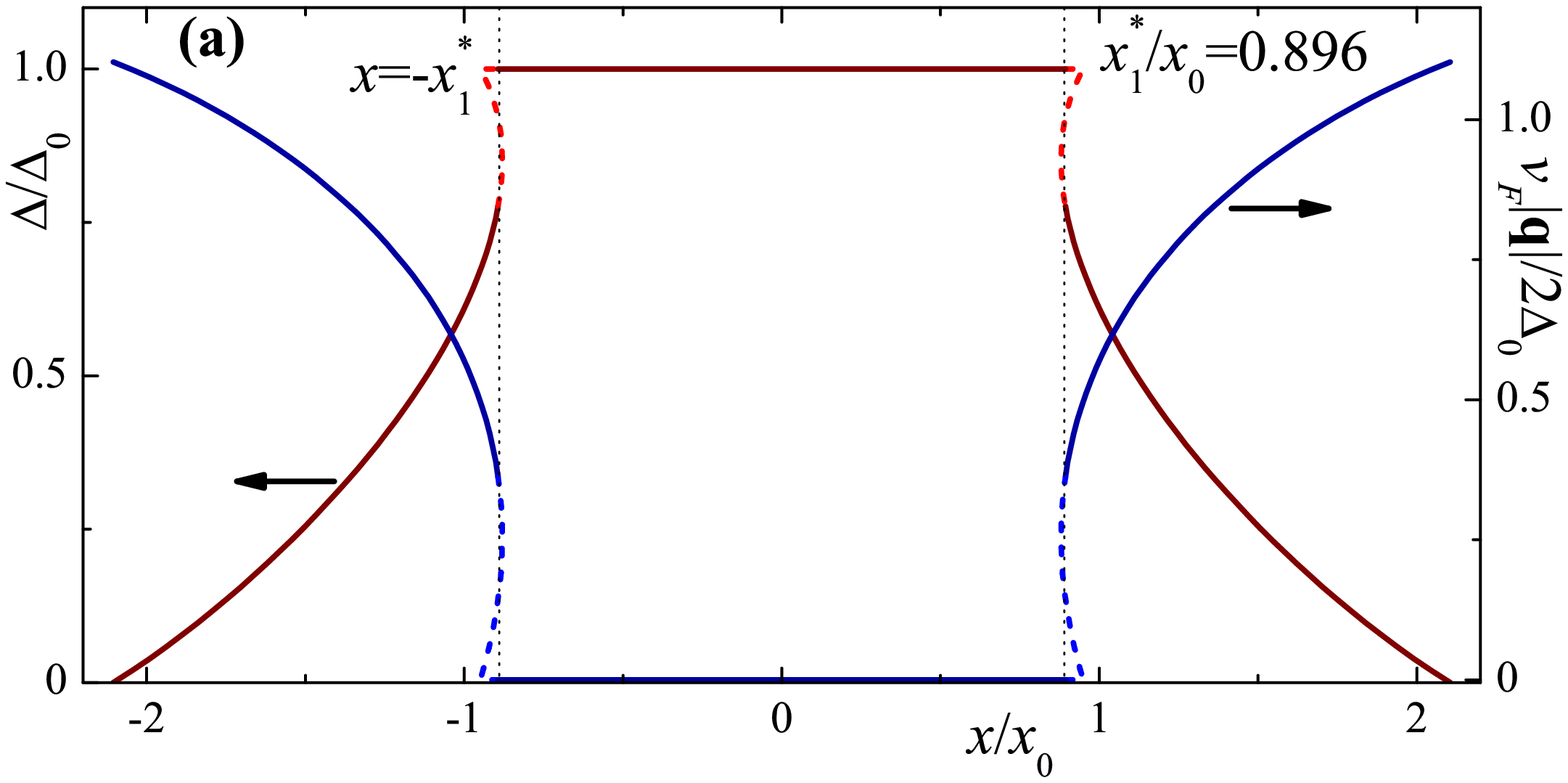}\\
    \includegraphics[width=0.98\columnwidth]{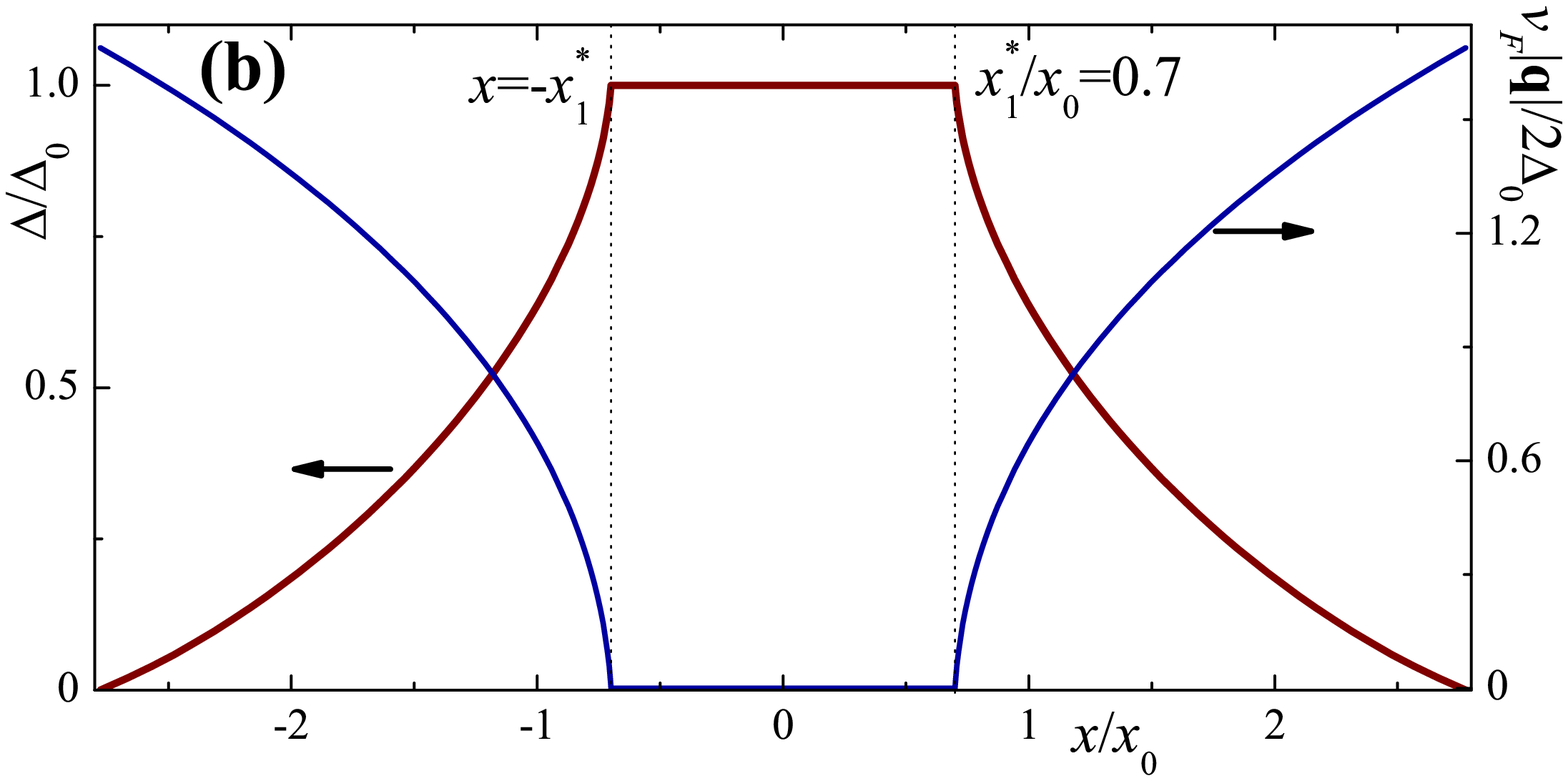}\vspace{6mm}\\
    \includegraphics[width=0.98\columnwidth]{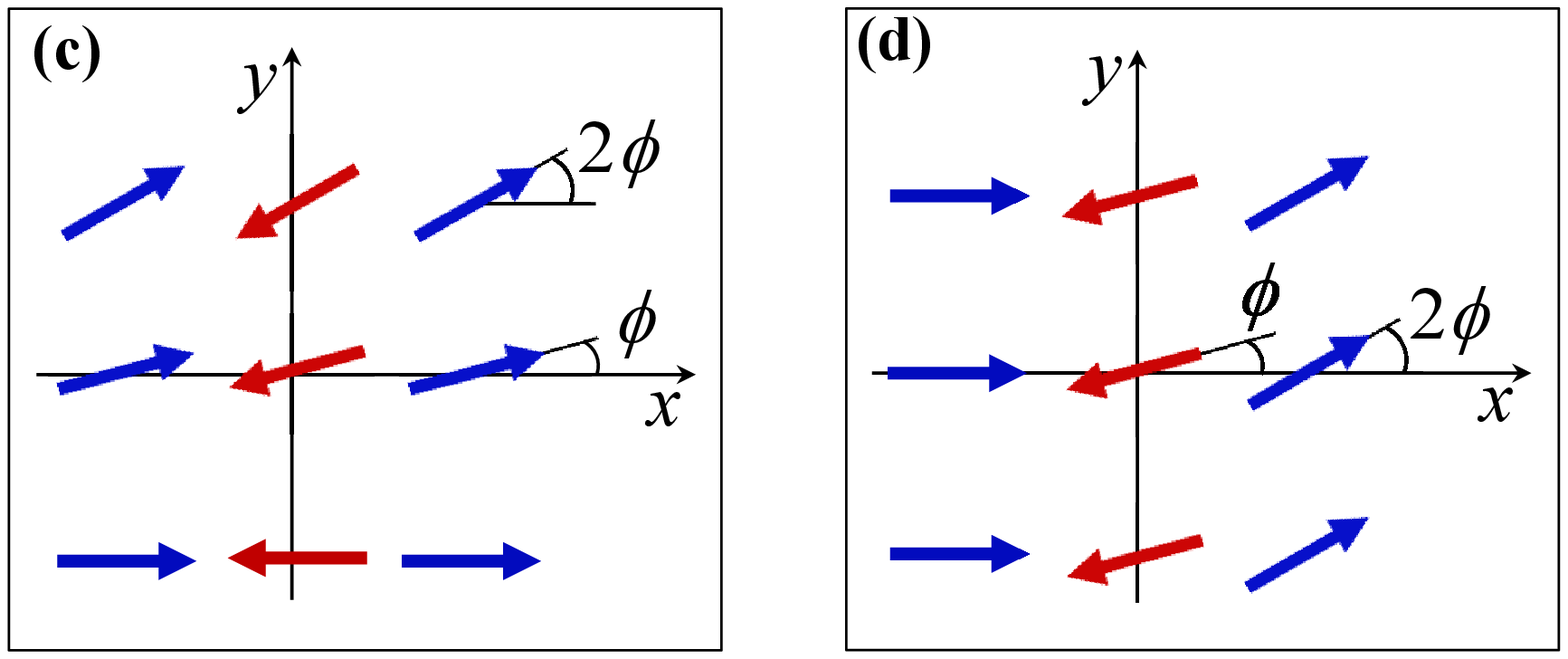}
 \caption{(Color online) Dependence of the normalized order parameter $\Delta/\Delta_0$ and wave vector $|\mathbf{q}|$
on the doping $x$, for $r=\varkappa=1$.
(a) Here $\bar{\alpha}=0.1$. The transition from commensurate to
incommensurate SDW state is of first order: both $\Delta$ and
$|\mathbf{q}|$ abruptly change at $x=x_1^{*}$. The dashed curves show the functions $\Delta(x)$ and $|\mathbf{q}(x)|$
corresponding to the metastable state. (b) Here $\bar{\alpha}=0.6$. The transition from commensurate to incommensurate SDW state is now of second order. Panels (c) and (d) show schematics of the incommensurate SDW spin structure. If $\alpha>0$, then panel (c) corresponds to hole doping ($x<0$) and (d) to electron doping
($x>0$). If $\alpha<0$, then panel (c) corresponds to electron
doping, while panel (d) corresponds to hole doping.}
\label{FigDeltaQ}
\end{figure}

The computed functions $\Delta(x)$ and $q(x)$ are shown in
Fig.~\ref{FigDeltaQ}(a) and (b) for $r=\varkappa=1$
and two different values of $\bar{\alpha}$.
At low doping, $|x|<x_1^{*}$, where the critical doping
$x_1^*$ depends on model parameters, all the extra charge goes to the nonmagnetic bands. As a result, the order parameter is independent of $x$, $\Delta(x)=\Delta_0$, $\mathbf{q}(x)=0$, and the chemical potential increases linearly with $x$,
\begin{eqnarray}
\mu(x)=\frac{\Delta_0 x}{rx_0}.
\end{eqnarray}
In such a regime, the system exhibits a commensurate SDW order.

When $|x|=x_1^*$, the chemical potential touches the bottom (top) of the upper (lower) magnetic band. Upon further doping, electrons (holes) appear in the band $E^{(2)}_{\mathbf{k}}$ ($E^{(1)}_{\mathbf{k}}$). The SDW order becomes incommensurate,~\cite{ic_sdw} ${\bf q} \ne 0$. The order of the transition into the incommensurate state depends on the
parameters $\bar{\alpha}$ and $r$. If $r<r_1\cong0.38$,
the transition is of second order for any $\bar{\alpha}$,
while for larger $r$ it becomes of first order if
$|\bar{\alpha}|<\bar{\alpha}_1(r)$. In the case of first-order transitions, both $\Delta$ and $|\mathbf{q}|$ abruptly change at
$x=x_1^{*}$, see Fig.~\ref{FigDeltaQ}(a).

Electron doping ($x>0$) is in many respects similar to hole doping ($x<0$). Indeed, the functions $\Delta(x)$, $\mu(x)$, and $|\mathbf{q}(x)|$ obey particle-hole symmetry relations
\begin{eqnarray}
\Delta(-x)=\Delta(x),
\\
\mu(-x)=-\mu(x),
\\
|\mathbf{q}(-x)|=|\mathbf{q}(x)|.
\end{eqnarray}
In addition, these functions are independent of the sign of $\alpha$. However, the direction of the vector $\mathbf{q}(x)$ depends on the sign of $\alpha$ and on the type of doping. When $\alpha>0$,
the vector $\mathbf{q}(x)$ is parallel to the $x$ axis for electron doping and to the $y$ axis for hole doping. The direction of
$\mathbf{q}(x)$ is reversed if $\alpha<0$. Thus, the nesting vector
$\mathbf{Q} = (Q_x, Q_y)$ is given by the following formulas
\begin{eqnarray}
\label{Qdir1}
Q_x &=& \frac{\pi}{a}
+
\frac{2\Delta_0p(|x|)}{\varkappa v_F},
\quad
Q_y = 0,\;\;
\mbox{if }
	\alpha x>0,
\\
\label{Qdir2}
Q_x &=& \frac{\pi}{a},
\quad
Q_y = \frac{2\Delta_0p(|x|)}{\varkappa v_F},\;\;
\mbox{if }
	\alpha x<0.
\end{eqnarray}
How does the incommensurate SDW described by these equations look like in real space? Let us introduce the integer-valued vector
$\mathbf{n}=(n,m)$. Then the position of a given iron atom is equal to ${\bf R}_{\bf n} = a {\bf n}$. For the order parameter $\Delta$,
Eq.~(\ref{DeltaQ}), the SDW magnetization vector lies in the $xy$ plane: $S^z_{\bf n} = 0$. The in-plane components can then be expressed as
\begin{eqnarray}
\mathbf{S_n}=S_0
\Big(
	\cos(a\mathbf{Q \cdot n}),\,
	\sin(a\mathbf{Q \cdot n})
\Big),
\end{eqnarray}
where
$S_0=\Delta/V_1$.
Thus, we obtain
\begin{equation}\label{Sdir}
\mathbf{S_n}=S_0\left\{
\begin{array}{rccl}
&\!\!(-1)^n(\cos\phi n\,,&\!\!\sin\phi n )\!\!&\,,\;\;\alpha x>0\,,\\
&\!\!(-1)^n(\cos\phi m\,,&\!\sin\phi m )\!\!&\,,\;\;\alpha x<0\,,
\end{array}\right.
\end{equation}
where $\phi=a|\mathbf{q}|$.
These spin configurations are schematically shown in Fig.~\ref{FigDeltaQ} (c) and (d) for hole and electron doping, respectively. For positive (negative) $\alpha$, panel~(c) corresponds to $x>0$ ($x<0$), and panel~(b) corresponds to $x<0$ ($x>0$).

\section{Phase separation}
\label{separation}

In the previous section we assumed that the ground state of the model is homogeneous. Here we demonstrate that there is a part of the phase diagram where homogeneous states are not stable, and the true ground state is phase separated. To detect this instability, the chemical potential must be calculated.

\begin{figure}[t]\centering
    \includegraphics[width=0.95\columnwidth]{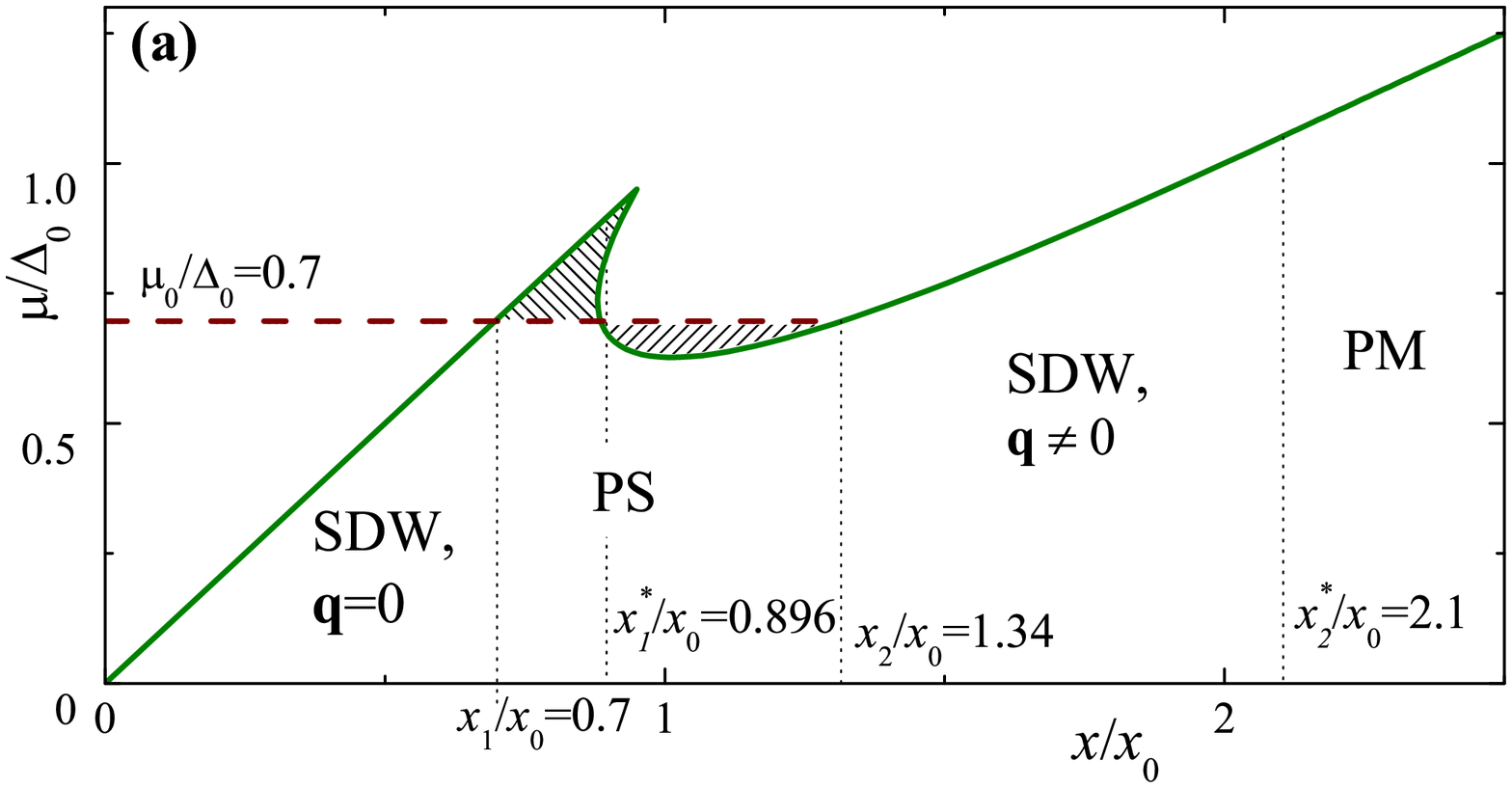}\\
    \includegraphics[width=0.95\columnwidth]{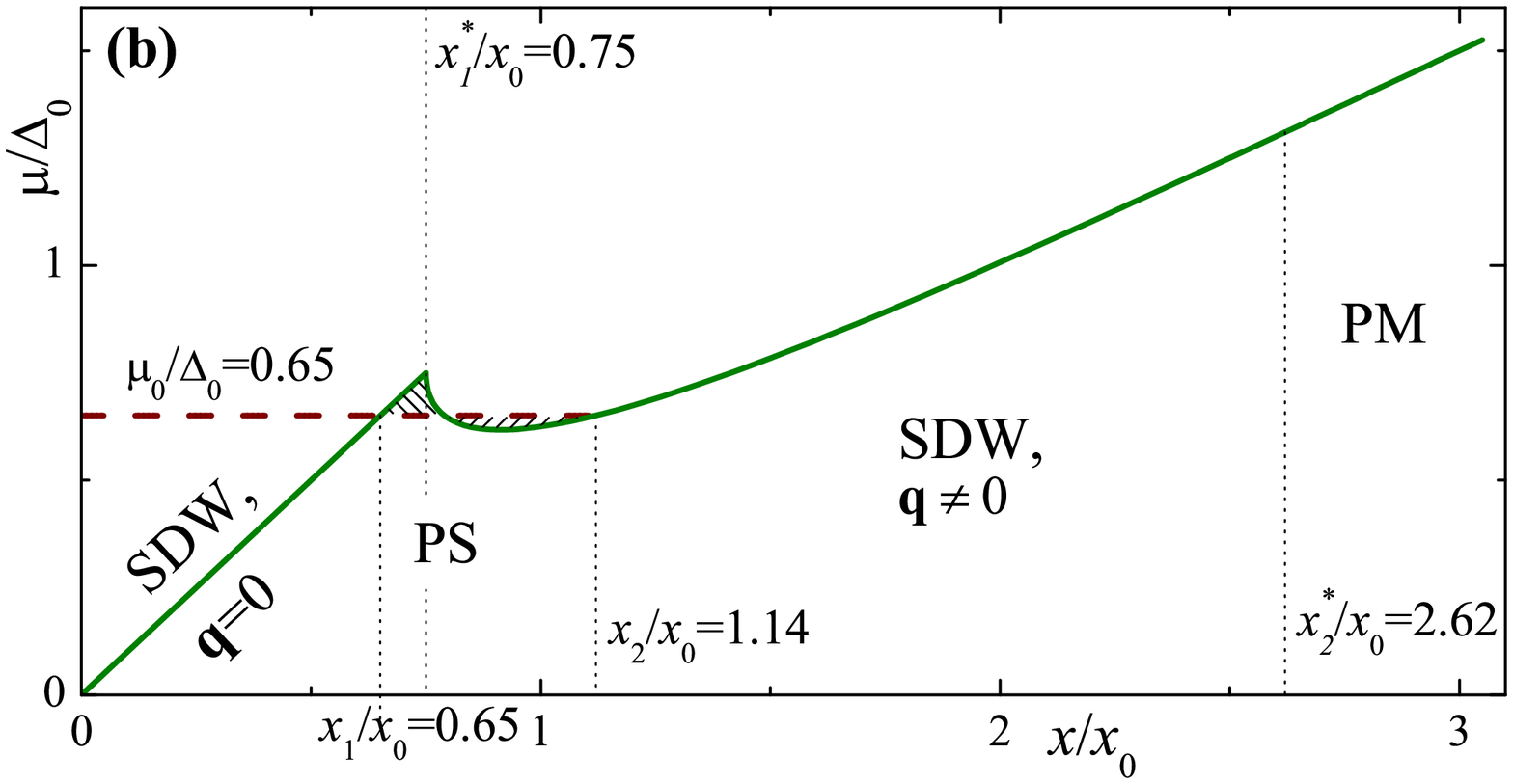}\\
    \includegraphics[width=0.95\columnwidth]{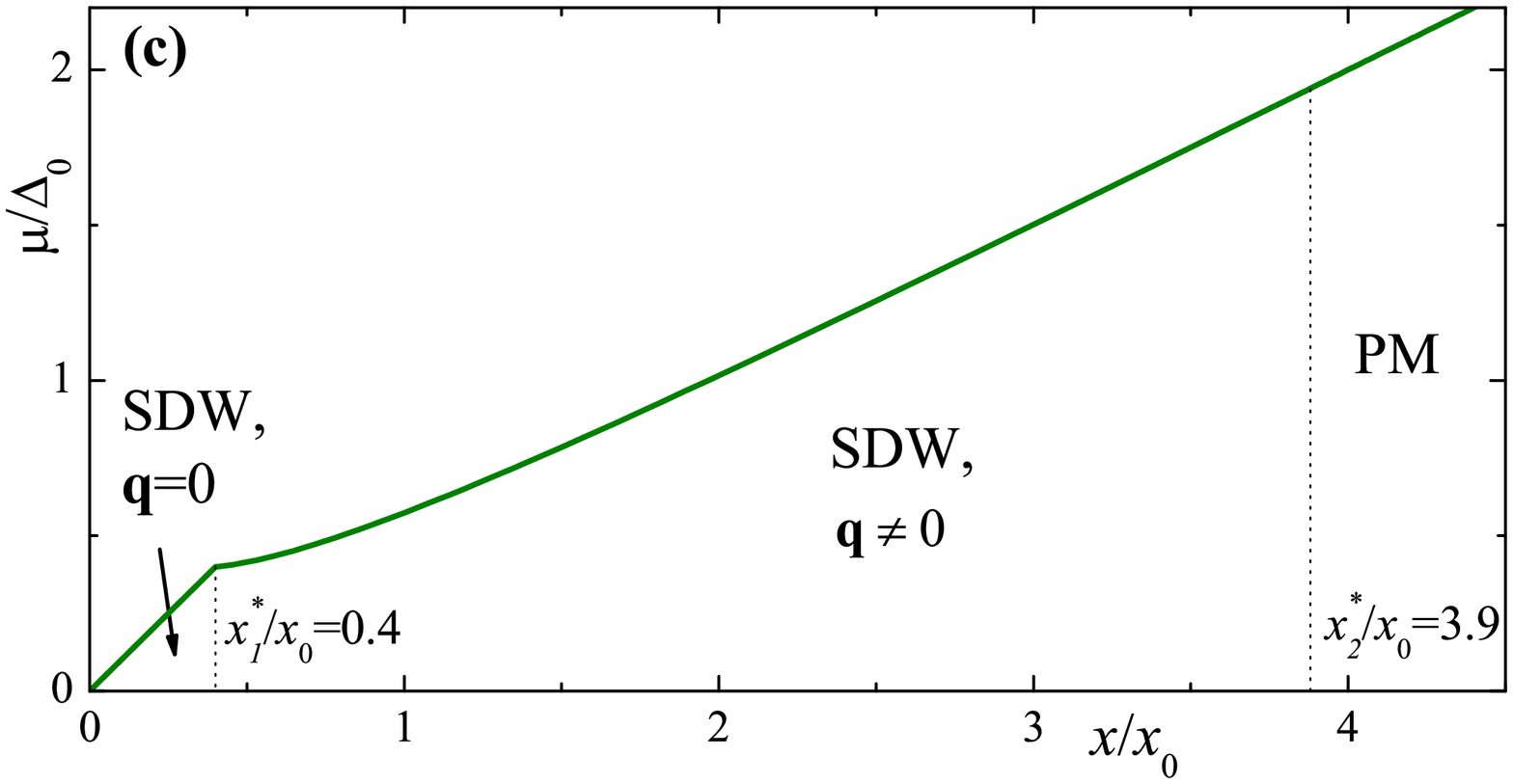}
 \caption{(Color online) Chemical potential $\mu(x)$ calculated at
$\bar{\alpha}=0.1$ (a), $\bar{\alpha}=0.5$ (b), and
$\bar{\alpha}=1.2$ (c). Here, $r=\varkappa=1$.
(a) and (b): the homogeneous state is {\it unstable} toward phase separation if $x_1<x<x_2$. The dashed (red) curve corresponds to the
$\mu_0$ found using the Maxwell construction. The shaded areas above and below $\mu_0$ are equal to each other. (c): $\mu(x)$ monotonically increases with $x$, no phase separation appears. The homogeneous commensurate ($\mathbf{q}=0$) and incommensurate ($\mathbf{q}\neq 0$) SDW, paramagnetic (PM), and inhomogeneous commensurate--incommensurate SDW (PS) states are separated by
vertical dotted lines.}\label{FigMu}
\end{figure}

The computed dependence of the chemical potential $\mu$ on doping $x$ is shown in Fig.~\ref{FigMu} for three different values of $\bar{\alpha}$ and $r=\varkappa=1$ for electron doping ($x>0$).
The dependence $\mu(x)$ for hole doping can be easily determined using particle-hole symmetry, as discussed in the previous section.

The function $\mu(x)$ shown in Fig.~\ref{FigMu} demonstrates several peculiar features in the vicinity of the commensurate-incommensurate phase transition point $x_1^{*}$. Specifically, for low $\bar{\alpha}$, the function $\mu(x)$ is non-monotonic and multivalued near $x \approx x_1^*$ [see Fig.~\ref{FigMu}(a)]. At higher $\bar{\alpha}$, the multivaluedness disappears; however, the non-monotonicity remains [see Fig.~\ref{FigMu}(b)]. This vanishes at even higher values of $\bar{\alpha}$ [as in Fig.~\ref{FigMu}(c)].

We now observe that in Fig.~\ref{FigMu}(a) and Fig.~\ref{FigMu}(b)
there are finite ranges of doping where the chemical potential decreases as the doping increases. This means that the compressibility of the electronic system is negative and the homogeneous state is {\it unstable} with respect to separation into two
phases.~\cite{thermodyn} The phase transitions between homogeneous commensurate and incommensurate SDW phases, which we described in
subsection~\ref{results}, will be masked, at least partially, by the phase separation.

In the separated state there exist two phases, phase~$1$ and phase~$2$, with electron density $x_1$($<x_1^*$) and $x_2$($>x_1^*$), and with the volume fractions $p_1$ and $p_2$ satisfying the conditions $p_1+p_2=1$ and $x_1p_1+x_2p_2=x$. As one can see from Fig.~\ref{FigMu} (a,b), the phase $1$ is the commensurate and the phase $2$ is the incommensurate SDW state. The concentrations $x_1$ and $x_2$ are found from the equations~\cite{thermodyn}  $\mu(x_1)=\mu(x_2)\equiv\mu_0$ and $\Omega_1=\Omega_2$, where $\Omega_{1,2}$ are grand potentials in the phases $1$ and $2$. The latter condition means the equality of two shaded areas shown in Fig.~\ref{FigMu}(a,b) (the so-called Maxwell construction).

The range of doping $x$, where the phase separation exists, is largest when $\alpha=0$. In this case, our model is identical to the two-dimensional Rice model~\cite{RicePRB70}, for which the presence of the phase separation was shown in Refs.~\onlinecite{tokatly,OurRicePRB}. The range of the phase separation, $x_1<x<x_2$, shrinks if $|\bar{\alpha}|$ increases, and disappears at the critical value $\bar{\alpha}_c\cong1.15$. The phase separated state does not exist for $|\bar{\alpha}|>\bar{\alpha}_c$.

\begin{figure}[t]\centering
    \includegraphics[width=0.95\columnwidth]{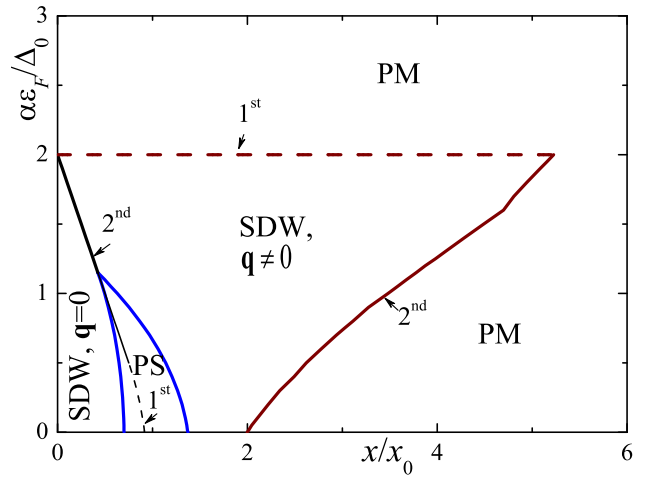}
\caption{(Color online) The phase diagram of the model~\eqref{Hmr} in the ($x,\bar{\alpha}$) plane; for $r=\varkappa=1$ and
$\alpha,x>0$. It is symmetric with respect to the replacement $x\to-x$ and/or $\alpha\to-\alpha$.  The boundary between incommensurate SDW and paramagnetic (PM) states shown by the dashed (red) curve corresponds to the first order and by the solid (red) curve to the second order phase transition. Solid (blue) lines indicate the boundaries of the phase separated (PS) state. The solid (black) curve (second order transition) and dashed (black) curve (first order transition) show the boundaries between commensurate ($\mathbf{q}=0$) and incommensurate ($\mathbf{q}\neq0$) homogeneous SDW phases. Note that phase separation partially masks the transition line between the homogeneous SDW states.
}
\label{FigPhDiag}
\end{figure}

The obtained results are summarized in the phase diagram in the
($x,\bar{\alpha}$) plane shown in Fig.~\ref{FigPhDiag}. This phase diagram is calculated for $r=\varkappa=1$. It remains qualitatively the same if $r\neq0$. If the nonmagnetic bands are absent, $r=0$, the homogeneous commensurate SDW phase exists only when $x=0$. Consequently, the electronic concentration $x_1$ in the phase 1 is zero in the phase separated state for any $\bar{\alpha}$.

Finally, we would like to draw the attention of the reader to an unexpected feature of the phase diagram in Fig.~\ref{FigPhDiag}.
Namely, the second-order incommensurate SDW-to-paramagnet transition line bends to the right, while all other transition curve in the phase diagram bend to the left. In other words, the value of doping where the transition from paramagnetic to incommensurate SDW phase occurs increases when the ellipticity parameter $\alpha$ increases. Interpreting this feature, however, one must keep in mind that the nesting quality is controlled not only by $\alpha$ but also by $\mu$ (or $x$). The parameter $\alpha$ controls the shape of the electron pocket, while $\mu$ (or doping) controls the relative areas of electron and hole pockets. Therefore, at larger doping levels, when the areas of the electron and hole pockets differ substantially, the electron and hole pockets could be better nested for larger ellipticity of the electron pocket.

\section{Discussion}
\label{discussion}

The most important result obtained here is the prediction of electronic phase separation in some range of doping. The separated phase consists of a mixture of commensurate and incommensurate SDW phases with different electronic concentrations. The phase separation in iron-based superconductors was observed in several experiments~\cite{PSexp1,PSexp2,PSexp3,PSexp4}. For example, the inhomogeneous state with a commensurate antiferromagnetic and nonmagnetic domains with characteristic sizes $\sim65$\,nm was
observed in the hole doped Ba$_{1-x}$K$_x$Fe$_2$As$_2$
compound.~\cite{PSexp1} Our theory predicts that the second phase is an incommensurate SDW rather than a nonmagnetic one. However, the proposed mechanism of phase separation can, in general, be consistent with the observations reported in Ref.~\onlinecite{PSexp1}.
We found that the thermodynamic potentials of the incommensurate SDW and the metastable paramagnetic phases are very close to each other in the doping range $x>x_2$. The incommensurate SDW phase can be destroyed by an additional reason not taken into account in our model, e.g., by disorder. In this case, the phase separation might occur between the commensurate SDW and the paramagnetic phases. In addition, the incommensurate SDW order parameter may be difficult to detect due to its weakness.

The geometry of the emergent inhomogeneities and their characteristic sizes are beyond the scope of the present study. The electron concentrations in separated phases are different; hence, the inhomogeneities are charged and one should take into consideration the electrostatic contribution to the total energy. The characteristic sizes of the inhomogeneities are controlled by the interplay between the long-range Coulomb interaction and the energy of the surface between the phases. In the simplest case, the
structure of the inhomogeneous state corresponds to the droplets of one phase embedded in the matrix of another phase. However, depending on the properties of the system, other geometries are possible, such as alternating layers of different phases, stripes,
etc.~\cite{Coulomb_refs} When disorder is present in the sample, it also affects the structure of the inhomogeneous phase.

In our study we chose the order parameter $\Delta$, Eq.~(\ref{DeltaQ}). In the literature, a different order parameter is also discussed.~\cite{tokatly,gorkov_teit_prb2010} This order parameter is not homogeneous: SDW gap experiences periodic
modulations in real space. When the state with such an order parameter is doped, the extra charges go to the places where the gap
locally vanishes.
The latter type of order may be more stable than Eq.~(\ref{DeltaQ}). However, for several reasons we decided to avoid such an option.
Specifically, analytical calculations with inhomogeneous gaps become very complicated. Further, the choice of the order parameter affects somewhat the phase diagram and, in particular, the region of the phase separation; however, the inhomogeneous region does not disappear. In addition, we must remember that the relative stability of different order parameters is likely a non-universal quantity, which depends on a variety of microscopic parameters (e.g., details of the band structure, interaction, disorder). Therefore, the type of suitable order parameter cannot be deduced without input from experiments.

It follows from the phase diagram in Fig.~\ref{FigPhDiag} that the smaller effective ellipticity parameter $\bar{\alpha}$, the larger the region of the phase separation. The SDW gap $\Delta_0$ increases with increasing coupling potential $V_1$. Therefore, a stronger coupling is more favorable for phase separation since $\bar{\alpha}\propto\alpha/\Delta_0$ [see Eq.~\eqref{alpha}].

Our mean-field approach is not applicable in the limit of strong coupling, where the use of Hubbard-like models is more appropriate. The phase separation in such models is a common
phenomenon~\cite{Dagbook,Nagbook,KaKuUFN2001,LittlwNature2005,
DagottoNJP2005,OvchinUFN2009,KugelPRL2005,DagottoScience2005,SboychaPRB2007}.
Thus, we expect that the existence of phase separation in our model is not an exclusive feature of the weak-coupling regime.

The model studied here predicts an incommensurate SDW phase for both
electron and hole doping in the range of dopant concentrations scaled by the parameter $x_0$, Eq.~\eqref{defs1}. From the latter equation we conclude that $x_0\ll 1$ if the weak-coupling condition, Eq.~(\ref{limitV}), is met.
Thus, we predict the existence of the incommensurate SDW order for small doping, $x\ll 1$.

The incommensurate SDW phase is characterized by the vector
$\mathbf{q}=\mathbf{Q}-\mathbf{Q}_0$. Our calculations show that there are only two possible equilibrium directions of $\mathbf{q}$: it can be either parallel, or perpendicular to $\mathbf{Q}_0$, depending on the type of doping and the sign of the ellipticity parameter $\alpha$ [see Eqs.~\eqref{Qdir1} and~\eqref{Qdir2}].
It is clear, however, that both the magnitude and direction of $\mathbf{q}$ is sensitive to the shape of the Fermi surface, which is simplified in the present calculations. In real materials, the shape of the hole pockets deviates from perfect circles. Moreover, the spectrum of the hole and electron bands depend on the transverse momentum, $k_z$, which is completely neglected in the model considered here. In particular, the electron bands have corrugated
structure~\cite{GraserEll1,BrouetEll2}, that is, the major axis of the ellipses are rotated by $90^{\circ}$ when $k_z$ varies from $0$ to $\pi$. Nevertheless, we believe that the model studied here captures the main features of the incommensurate SDW state in pnictides.

Indeed, the observation of the incommensurate SDW phase with $\mathbf{q}$ perpendicular to $\mathbf{Q}_0$ in the electron doped
Ba(Fe$_{1-x}$Co$_x$)$_2$As$_2$ was recently reported in Ref.~\onlinecite{PrattICSDWed}. The measured value of $|\mathbf{q}|$ was about $0.02$--$0.03$ for the concentration range $0.056<x<0.06$.
Such values of $x$ and $\mathbf{q}$, as well as the direction of
$\mathbf{q}$ (corresponding to $\alpha<0$), are consistent with our theory.

The short-range incommensurate SDW phase was described in
Ref.~\onlinecite{LuoSRICed} for the electron doped
BaFe$_{2-x}$Ni$_x$As$_2$. In general, one must be cautious interpreting this experiment using our model, since the observed short-range SDW correlations do not correspond to our long-range SDW order. However, we note that the measured vector $\mathbf{q}$
was perpendicular to $\mathbf{Q}_0$, which is also consistent with a negative $\alpha$.

We found no experimental work reporting the observation of the
incommensurate SDW phase in the hole-doped pnictides. However,
double-peaked spin fluctuations at $\mathbf{Q}=\mathbf{Q}_0\pm\mathbf{q}$
have been observed~\cite{LeeICSFhd,CastellanICSEhd}
in the hole-doped Ba$_{1-x}$K$_x$Fe$_2$As$_2$
compound in a wide doping range. In these
measurements~\cite{LeeICSFhd,CastellanICSEhd}
the vector $\mathbf{q}$ was found to be parallel to
$\mathbf{Q}_0$, which, again, corresponds to $\alpha<0$ in our model.

Note that, in general, all electron and hole parts of the Fermi surface interact with each other. Each $i$th pair of these interacting electron-hole surfaces are characterized by the nesting vector $Q_{i}$, coupling constant $V_i$, and ``de-nesting'' parameter $\alpha_i$. However, the $i$th interaction contributes to the SDW order only if $V_i$ exceeds some critical value (see Section~\ref{results}). All other charge carriers do not contribute to the ``magnetic'' interaction and form a reservoir. It is rather quite probable that among the various iron pnictides there exist systems with such reservoirs. However, the existence of the reservoir is not necessary for the nucleation of the incommensurate AFM order or for phase separation.

We limit our consideration to only one type of interaction. Even this simple case corresponds to the experimentally-observed SDW symmetry both for commensurate and incommensurate phases. Also, ``switching on'' additional interactions may be necessary in the future, if the incommensurate AFM ordering with different symmetry would be observed. However, let us emphasize that the qualitative results of our work remain unchanged.

\section{Conclusions} \label{Concl}

We considered a model with imperfect nesting of the Fermi surface suitable for the description of iron pnictides. We show that an incommensurate SDW phase arises at finite doping. We demonstrate that two spin configurations can arise in the system depending on the model parameters. It was shown that the homogeneous state is unstable toward phase separation into commensurate and incommensurate SDW phases in a specific doping range. These results are in qualitative agreement with the recent experimental observation of the incommensurate SDW order and phase separation in doped superconducting pnictides.

The main achievement of this paper is the demonstration that a simple model approximating the Fermi surface of pnictides implies the existence of electronic phase separation even in the weak-coupling regime, that is, in the absence of strong electron correlations. This is an important finding for the interpretation of the experimental data on phase inhomogeneity of iron pnictides: it proves that a purely electronic model with moderate interaction is sufficient to explain the observed inhomogeneities.

There are several features of pnictides which we incorporate into our approach. Let us briefly recall them.

First, we need a Fermi surface with sufficient nesting (poor nesting implies no phase separation). The ARPES data~\cite{RichardRPrPh201}support the assumption that at least some sheets of the Fermi surface nest adequately.

Second, the interaction between electrons in the nested bands must not be too weak: the ground state of the parent compound should be the homogeneous SDW state. There is plenty of experimental evidence that a broad class of iron pnictides are in such a state at zero doping. However, to make our Hamiltonian analytically tractable, we discarded all interaction constants except one, $V_1$. Further, $V_1$ was assumed to be momentum-independent.

Third, following Rice~\cite{RicePRB70}, we included non-magnetic (``reservoir") bands into our model. This is a simple way to account for those sheets of the Fermi surface which do not nest with some other sheets. The Fermi surface pockets, which may serve as the reservoir, are seen by ARPES (like ``propellers'', see, e.g.  Ref.~\onlinecite{BorisenkoJPSJ2011}). In addition, the finite conductivity of pnictides even in the undoped SDW state may be easily explained by non-magnetic bands (although, other explanations are also possible).

Clearly, our model, being quite simple and generic, cannot predict whether a particular material exhibits phase separation. One can attempt devising a more elaborate model. For example, instead of a single coupling constant $V_1$, a set of spin-dependent couplings
$V_{\alpha \beta}^{\sigma \sigma'}$, which describe the interaction of electrons in the band $\alpha$ and the band $\beta$, may be introduced. Other refinements, e.g., 3D structure of the material, orbital content of the Fermi surface states~\cite{ScalapinoPRB2010,Suzuki2011} (the latter introduces a pronounced momentum dependence of the interaction), are possible as well.

Yet, such a detailed investigation can be counterproductive. Indeed, our analysis clearly shows that the phase separation is a non-universal feature of the model. That is, its presence depends on the parameters of the Hamiltonian. It is very likely that a study of a a more complex model, while being quite costly in terms of effort, will bring up the same conclusion about the non-universality of phase separation. Thus, in the situation where accurate knowledge of the microscopic parameters is absent, one would not be able to reliably establish the presence of phase separation for a given material. Keeping these circumstances in mind, we emphasize that the main purpose of our study is to introduce the mechanism of phase separation in iron pnictides. However, the problem regarding the instability of the homogeneous state for a particular material has to be solved on an individual basis by the analysis of the specific feature of its electronic structure.

\section*{Acknowledgments}

This work was partially supported by the ARO, RIKEN iTHES Project,
MURI Center for Dynamic Magneto-Optics, JSPS-RFBR contract no. 12-02-92100, Grant-in-Aid for Scientific Research (S), MEXT Kakenhi on Quantum Cybernetics, the JSPS via its FIRST program, and by the Russian Foundation for Basic Research (projects 11-02-00708, 11-02-00741, and 12-02-00339). A.O.S. acknowledges support from the RFBR project 12-02-31400 and the Dynasty Foundation.

\end{document}